\documentclass[twocolumn,english,pre]{revtex4}
\usepackage{ae}
\usepackage{aecompl}
\usepackage[T1]{fontenc}
\usepackage[latin1]{inputenc}
\usepackage{float}
\usepackage{graphicx}

\makeatletter

\newcommand{\lyxdot}{.}

\floatstyle{ruled}
\newfloat{algorithm}{tbp}{loa}
\floatname{algorithm}{Algorithm}

\usepackage{babel}
\makeatother
\begin{document}

\title{Hysteretic Optimization For Spin Glasses}

\author{B. Gonçalves and S. Boettcher}

\affiliation{Emory University, Atlanta, Ga 30322}

\email{bgoncalves@physics.emory.edu}

\begin{abstract}
The recently proposed Hysteretic Optimization (HO) procedure is applied
to the $1D$ Ising spin chain with long range interactions. To study
its  effectiveness  , the quality of ground state energies
found as a function of the distance dependence exponent, $\sigma$,
is assessed. It is found that the transition from an infinite-range
to a long-range  interaction at $\sigma=0.5$ is accompanied by a
sharp decrease in the performance     . The transition
is signaled by a change in the scaling behavior of the average avalanche
size observed during the hysteresis process. This indicates that HO
requires the system to be infinite-range, with a high degree of interconnectivity
between variables leading to  large avalanches, in order to function
properly. An analysis of the way auto-correlations evolve during the
optimization procedure confirm that the search of phase space is less
efficient, with the system becoming effectively stuck in suboptimal
configurations much earlier. These observations explain the poor performance
that HO obtained for the Edwards-Anderson spin glass on finite-dimensional
lattices, and suggest that its  usefulness  might be limited
in many combinatorial optimization problems.
\end{abstract}
\maketitle

\section{Introduction}

The steadily increasing availability of powerful computational resources
has allowed scientists and engineers alike to study ever more realistic
and complex problems. Historically, a significant fraction of all
available computer time has been used to traverse phase space in search
of the optimal solution to a given problem. 

Particularly, in several physics domains\cite{Dagstuhl04}, such as
spin glasses, disordered materials and protein folding, one is interested
in enumerating a large number of local minima of the energy landscape
\cite{Frauenfelder96,Wales03}, as that provides us with valuable
information about its physical properties.  Many  algorithms and heuristics
have been developed over the years to tackle this kind of problem.
In this article we study a recently proposed algorithm known as Hysteretic
Optimization (HO)\cite{zarand02,Pal03,Pal06,Pal06b}.  HO is motivated
by the physics of demagnetizing magnetic materials with a slowly oscillating
external field of decreasing amplitude. Similar to the thermal fluctuations
in simulated annealing \cite{SA} or the activated dynamics of extremal
optimization (EO) \cite{Boettcher01a}, the drag created by the external
field carries HO over energetic barriers. 

HO has proved efficient\cite{Pal06} at finding the ground state
configurations of the well know Sherrington-Kirkpatrick (SK)\cite{Sherrington75}
mean-field spin glass and of the classical Traveling Salesman Problem,
but  has, hitherto, been unsuccessful in searching  the corresponding
Edwards-Anderson spin glass on finite-dimensional lattices\cite{Edwards75},
as described in Chap. 10 of Ref. \cite{Dagstuhl04}. Similarly, it
was shown in Ref. \cite{Zapperi04} that HO performs poorly for the
random field Ising model (RFIM) on a one or three dimensional lattice.
(The RFIM is a classic model for disordered magnetic materials, but
unlike the spin glass case there are polynomial-time algorithms to
find global optima in the energy landscape, see Chap. 5 in Ref. \cite{Dagstuhl04}.)
Since efficient heuristics for hard (i. e. beyond polynomial) optimization
problems are still few and far between, especially for spin glasses,
but also for many other combinatorial problems\cite{Percus06a}, promising
new algorithms warrant careful investigation. Here, we explore the
behavior of HO under variation of the search space characteristics
that is representative of many problems. In particular, we apply HO
to a one-parameter family of spin glass problems \cite{Kotliar83,fisher85-1,fisher88-1,katzgraber03-1,katzgraber05-1}
that interpolates between the characteristics of the SK model on one
extreme and the EA model on the other. We observe that the break-down
of HO for the EA model is intimately linked to the physics of intermittent
events (i. e. avalanches) kicked-off by the external field in the
hysteresis loop. We find a distinct cross-over between broadly-distributed
avalanching dynamics in the SK-regime\cite{Pazmandi99}, connected
with a high degree of interconnectivity between variables and divergent
energy scales, and sharply cut-off dynamics in the EA regime. Unfortunately,
the need for strong interconnectivity between variables severely limits
the applicability of HO with respect to combinatorial problems related
to spin glasses of low degree such as satisfiability, partitioning,
or coloring at their respective phase transitions \cite{Percus06a}. 

This article is structured as follows, in Section II, we introduce
the general Hysteretic Optimization procedure and apply it to a generalization
spin glass  model. In Section III, we investigate in detail the avalanche
dynamics during the hysteresis process to identify the reasons that
lead to the breakdown of HO's performance.

\section{Hysteretic Optimization}

For a magnetic material, such as an Ising system, to obtain a zero
magnetization value, an a.c. demagnetization is performed. The sample
is placed in an oscillating and slowly decaying magnetic field. As
the amplitude of the external field approaches zero, so does the magnetization.
At low enough temperature and slow driving,   a disordered systems
gets dragged through a sequence of local energy minima. Based on this
observation, Zarand \emph{et al} \cite{zarand02} proposed Hysteretic
Optimization (HO) as a general-purpose local search heuristic\cite{Hoos04}
to explore the phase space of many combinatorial optimization problems.
As an example, their study implemented the HO algorithm as listed
in Tab. \ref{alg:HO}

          Our study
here is focused only on finding the ground state ($T=0$) energies
of spin glasses, for which case we describe the implementation of
HO in detail.

\begin{figure}
\begin{centering}
\includegraphics[clip,width=7cm]{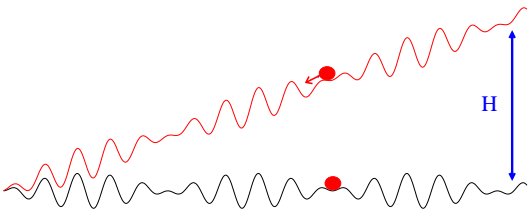}
\par\end{centering}

\caption{\label{fig:HO-Landscape}Simple sketch of a one-dimensional slice
through the energy landscape for a generic optimization problem as
in Eq. (\ref{eq:Hamiltonian}) without external field (bottom) and
in Eq. (\ref{eq:HamiltonianHO}) with external field (top).}
\end{figure}

In an Ising spin glass, each spin $S_{i}\in\left\{ \pm1\right\} $
is assumed to have a random bond $J_{i,j}$ with other spins $S_{j}$:\begin{equation}
\mathcal{H}=-\sum_{\langle i,j\rangle}J_{ij}S_{i}S_{j},\label{eq:Hamiltonian}\end{equation}
where the summation is taken over all pairs of spins. To find ground
states of this spin glass with HO, we   couple each spin $\sigma_{i}$
to an  external field of amplitude $H$.     
      with a random sign $\xi_{i}\in\left\{ \pm1\right\} $,
which may be adjusted even during a single demagnetization run. The
Hamiltonian  of this extended system is then:\begin{equation}
\mathcal{H}_{HO}=\mathcal{H}+H\sum_{i}\xi_{i}S_{i}\label{eq:HamiltonianHO}\end{equation}
 Physically, the second term in Eq. (\ref{eq:HamiltonianHO}) distorts
the energy landscape as shown in Fig. \ref{fig:HO-Landscape}, allowing
the system to escape  local minima. Fluctuations due to the coupling
to the external field can compensate for unsatisfied bonds.  By varying
the external field $H$ and the random couplings $\xi_{i}$, HO can
force the system to explore a vast area of phase space in search of
the optimal solution.

\begin{figure}
\begin{centering}
\includegraphics[clip,width=7cm,keepaspectratio]{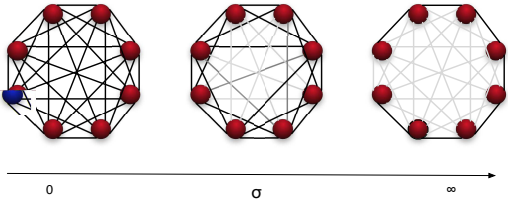}
\par\end{centering}

\caption{\label{fig:SKKY-alpha}Systems described by the long-range spin glass
model as a function of $\sigma$. For $\sigma\equiv0$ we re-obtain
the \emph{SK} model and as $\sigma\to\infty$ all the long range links
become essentially negligible leaving us only with nearest-neighbor
interactions.}
\end{figure}

\begin{algorithm}[H]

\caption{\label{alg:HO}Hysteretic Optimization}

\begin{enumerate}
\item Set  $H=H_{1}$ large enough  such that $S_{i}=\xi_{i}$ $\forall$
$i$. Set $E_{\{{\rm min\}}}=\mathcal{H}\left(=\mathcal{H}_{HO}\vert_{H=0}\right)$.
\item Decrease  $H$ until one spin becomes unstable and allow the system
to relax. If $\mathcal{H}<E_{\{{\rm min\}}}$, set $E_{\{{\rm min\}}}=\mathcal{H}$.
\item Optional: When $H$ passes zero,  randomize $\xi_{i}$,  leaving the
current configuration stable. 
\item At each turning point $H=H_{n}=-\gamma_{n-1}H_{n-1}$, for $0<\gamma_{n}<1$,
reverse the direction of $H$.
\item Terminate when amplitude  $|H_{n}|<H_{\{{\rm min\}}}$,.
\item Restart at 1 for $N_{run}$ times with a new, random set of $\xi_{i}$'s
.
\item Return the best $E_{\{{\rm min\}}}$ over all runs.
\end{enumerate}

\end{algorithm}

Following the prescription of Algorithm \ref{alg:HO} for the variation
of the external field $H$, each run, in effect, starts by exploring
a large region of phase space which subsequently  decreases slowly.
 By varying the field between positive and negative amplitudes $H_{n}=-\gamma_{n-1}H_{n-1}$,
  the runs repeatedly quench the system, following an approach similar
to the well known simulated annealing or tempering  algorithms\cite{SA,Salamon02,Marinari92}.

HO operates at $T=0$ , thus there are no thermal fluctuations and
we can simply calculate the field necessary to make the next spin
unstable and increase it to that value  (within the $\gamma^{n}H_{0}$
limit) . Typically, HO is run with multiple restarts from the largest
amplitude to increase the chances of finding a better approximation
to the global minimum. Note, however, that each run itself has no
stochastic element once the couplings $\xi_{i}$ to the external field
are fixed. It is therefore useful to restart the demagnetization process
repeatedly with a fresh set of random field directions (see
item 6. in Algorithm \ref{alg:HO}). In fact, it is also possible
to refresh the $\xi_{i}$ each time the external field $H$ passes
through zero during each run (see item $3$ in Alg. \ref{alg:HO}). 

This algorithm has been very successful in determining the ground
state energies of the Sherrington-Kirkpatrick spin glass\cite{Pal06}
and reasonably efficient for the Traveling Salesman Problem\cite{pal06-2},
but there are few attempts  to apply it to other problems\cite{Zapperi04}.
In this article we focus on a Ising spin glass on a one-dimensional
ring with power-law interactions\cite{Kotliar83,fisher85-1,fisher88-1,katzgraber03-1,katzgraber05-1}
defined by the Hamiltonian in Eq. (\ref{eq:HamiltonianHO}) with bonds
of the form: \begin{equation}
J_{i,j}=\frac{\epsilon_{ij}}{r_{ij}^{\sigma}}.\label{eq:sigmaBonds}\end{equation}
Here, $\epsilon_{ij}$ are random variables drawn independently from
a Gaussian distribution of zero mean and unit variance and\[
r_{ij}=\frac{L}{\pi}\sin\left(\frac{\pi\left|i-j\right|}{L}\right)\]
is the distance between each pair of spins on the ring. By varying
$\sigma$ we can interpolate between the all-to-all $SK$ limit ($\sigma=0$)
and the nearest-neighbor $EA$ limit (large $\sigma)$, as shown in
Fig. \ref{fig:SKKY-alpha}. This model has been extremely useful in
elucidating the connection between mean-field and finite-dimensional
spin glasses\cite{katzgraber05-1,katzgraber03-1}. 

\begin{figure}
\begin{centering}
\includegraphics[width=7cm]{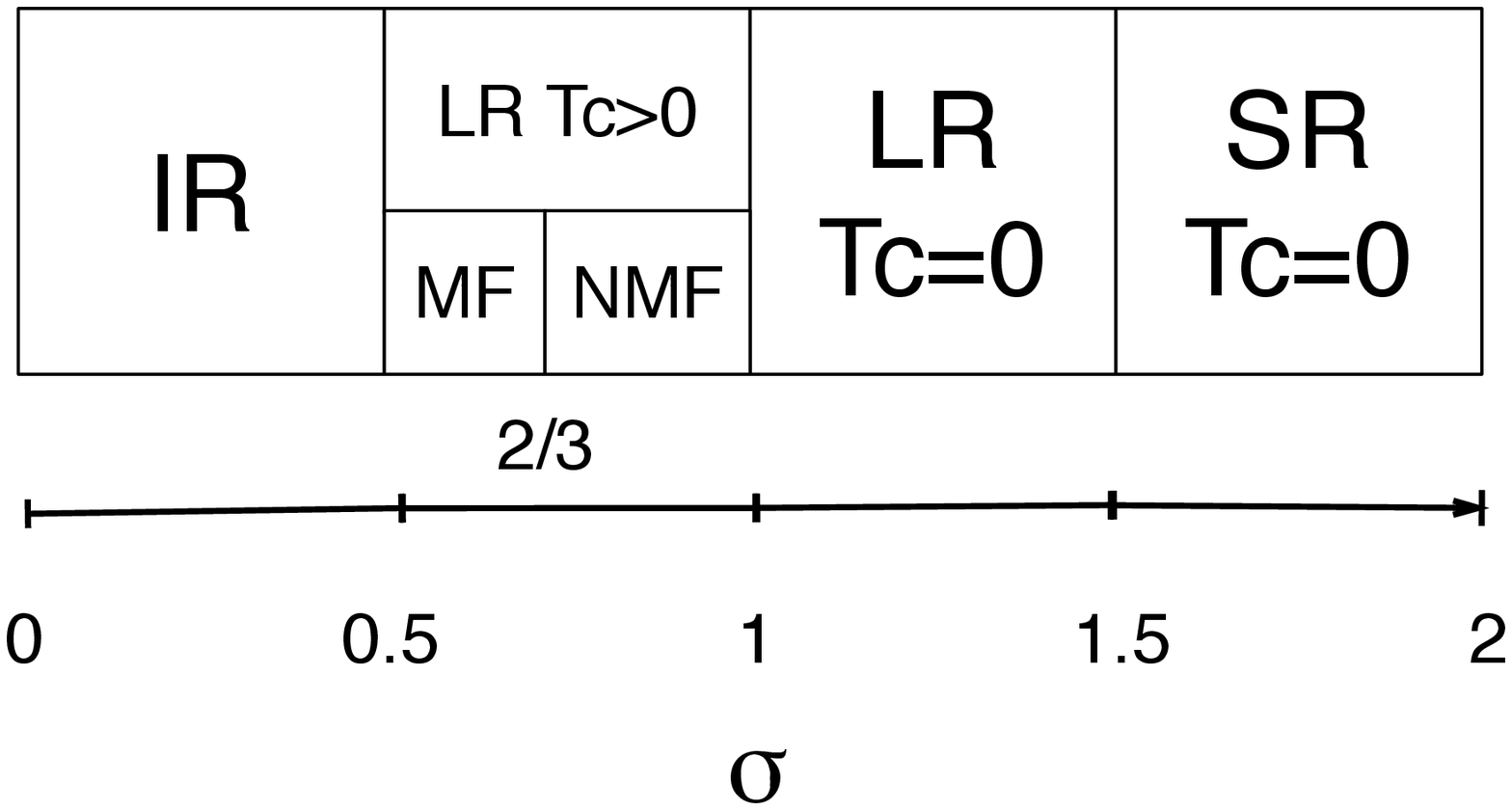}
\par\end{centering}

\caption{\label{fig:phase}Phase diagram, after \cite{Kotliar83,fisher85-1,fisher88-1,katzgraber03-1,katzgraber05-1}.}
\end{figure}

As it has been shown in the literature \cite{Kotliar83,Bray86,Fisher88},
as $\sigma$ is increased this spin glass goes through several distinct
phases, see Fig. \ref{fig:phase}. For $0\le\sigma<0.5$ the system
is effectively Infinite Range (IR). For all $\sigma$, the singular
part of the mean field transition temperature, $T_{c}^{MF}$, is of
the order \cite{katzgraber03-1}.:\[
\left(T_{c}^{MF}\right)^{2}\propto\sum_{i=2}^{N}\left[J_{i1}^{2}\right]_{av}=\sum_{i=2}^{N}r_{i1}^{-2\sigma}\sim N^{-2\sigma+1}\]
where $\left[\cdot\right]_{av}$ denotes an average over disorder
with   $[\epsilon_{ij}^{2}]_{av}=1$. This temperature becomes
finite in the thermodynamic limit at $\sigma=0.5$, signaling a transition
to a Long Range (LR) regime, where each node is able to see only a
finite fraction of the rest of the system. At $\sigma=1.0$, $T_{c}$
becomes zero    , but the LR character is preserved
until $\sigma=1.5$. From this point on, the structure of the system
is purely Short Range (SR), and each spin is connected only to $O\left(1\right)$
neighbors  

\begin{figure}
\begin{centering}
\includegraphics[clip,width=7cm,keepaspectratio]{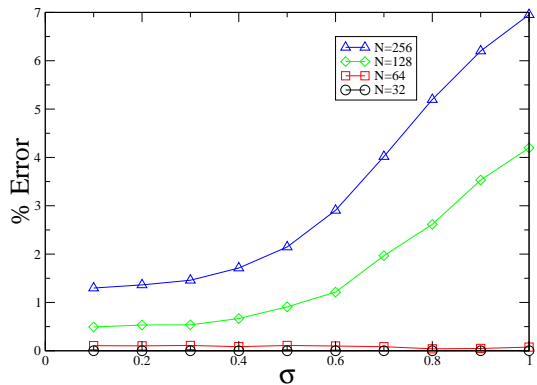}
\par\end{centering}

\caption{\label{fig:EHO}   Percentage difference between
the ground state found  by Hysteretic Optimization and the
one found by Extremal Optimization.}
\end{figure}

We have performed a benchmark study of the performance of $HO$ on
this spin glass model over a range of $\sigma-$ values. The point
of this study is not so much to tweak $HO$ for optimal performance,
but to obtain a clear assessment of its behavior under variation of
this parameter. To this end, we generated a benchmark of instances
of system sizes $N=32,$ $64$, $128$, and $256$, for which we have
obtained extremely good approximation to the ground state energy by
alternate means. In this case, we have used the Extremal Optimization
heuristic ($EO$) \cite{Boettcher00,boettcher05-1,Boettcher01a} but
expanding a large amount of CPU time to ensure accuracy. In fact,
using the implementation described in Ref. \cite{Boettcher01a}, $EO$
has proven itself equally capable of approximating ground states in
the SK model\cite{boettcher05-1} as for the EA\cite{Boettcher01a},
and it appears to be much less dependent on $\sigma$. Although a
direct comparison is not justified here due to the disproportionate
run times used for $EO$, we have found that even at much more extensive
runs, $HO$ was not able to find the exact-known ground state of a
one-dimensional spin glass with more than $\approx10^{2}$ spins\cite{Zapperi04}.)
In this set of instances, we have applied $HO$ with a minimal set
of control parameters. We set $\gamma=0.99$ and for each instance
in our set, we performed $10$ different runs, each with a separate
sets of $\xi_{i}$ that were kept constant throughout the entire run.
The ground state energy  was taken to be the best value seen over
10 different    quenches. This value was then averaged
over $1000$ different instances and compared with the results obtained
by EO for \emph{exactly} the same set of instances.

In Fig. \ref{fig:EHO} we plot the quality of the solution obtained
by $HO$ as a function of $\sigma$. The {}``Error'' is defined
as the percentage difference in ground state energy of the solutions
found using $HO$  relative to  $EO$. Generally, the quality of the
results found by $HO$ diminishes for increasing system size for all
$\sigma$, as can be expected with the limited CPU time (linear in
$N$) apportioned to these runs. More noticeable is the ever more
pronounced rise in error  for $\sigma>0.5$. To  understand the physical
reasons behind this behavior, we proceed to studying the dynamics
of the system in the next section.

\section{Avalanches And Correlations}

\begin{figure}
\begin{centering}
\includegraphics[clip,width=7cm]{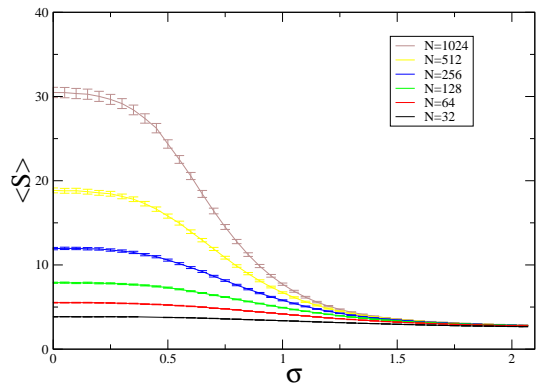}
\par\end{centering}

\caption{\label{fig:avgT}Average avalanche size as a function of $\sigma$
for different system sizes. }
\end{figure}

Unlike the comparison in the previous section, we now focus exclusively
on the intrinsic behavior of $HO$ itself. We will pinpoint  the causes
of $HO$s breakdown using quantitative measures. Using essentially
the same program as previously, for each value of $N$, we perform
$2$ (undamped) hysteresis cycles each, but for a much larger set
of $4\cdot10^{4}\cdot\sqrt{32/N}$ instances. Throughout, we set $\xi\equiv1$.

When the hysteresis procedure described causes a spin to become unstable
and get flipped, this may cause several other spins to become unstable,
thus initiating an avalanche in the system. Each avalanche can involve
a significant fraction of the number of spins in the system, including,
on occasion, several flips of the same spin in a form of long-range
self interaction. Avalanches have a wide range of sizes and can, in
principle, be larger than the system size by flipping the same spin
multiple times.

As a first step in our analysis, we measure $\left\langle S\left(\sigma\right)\right\rangle $,
the average avalanche size as a function of $\sigma$ at different
system sizes. This measurement will help us determine what range of
$\sigma$ we need to study, since it should become system size independent
in the nearest-neighbor limit. As we show on the right hand side
of Fig. \ref{fig:avgT}, this happens near $\sigma\approx2.0$, thus
restricting our interval of interest to $\sigma\in\left[0,2\right]$,
as expected from the literature\cite{katzgraber03-1}.

\begin{figure}[b]
\begin{centering}
\includegraphics[clip,width=7cm]{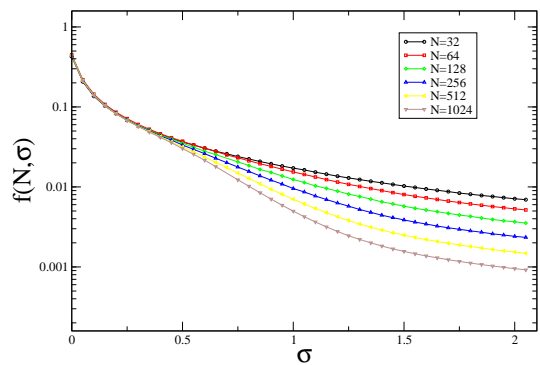}
\par\end{centering}

\caption{\label{fig:avgTscale}Corrections to scaling of the average avalanche
size, rescaled according to Eq. (\ref{eq:avalancheS}).}
\end{figure}

We find that this quantity obeys an empirical scaling relation  of
the form: \begin{equation}
\left\langle S\left(N,\sigma\right)\right\rangle \sim\frac{NA\left(\sigma\right)f\left(N,\sigma\right)+B\left(\sigma\right)}{\log^{2}\left(N\right)}\label{eq:avalancheS}\end{equation}
where $A\left(\sigma\right)$, $B\left(\sigma\right)$ are linear
functions of $\sigma$, and $f\left(N,\sigma\right)$ is plotted  in
Fig. \ref{fig:avgTscale}. The overlap of all the curves in the interval
$\sigma\in\left[0,0.5\right]$ means that all the $N$ dependence
has been captured by the $N/\log^{2}\left(N\right)$ term. This scaling
should be compared with the $N/\log(N)$ scaling found for this behavior
for $\sigma=0$ , i. e. the SK model, by Ref. \cite{Pazmandi99}.
In fact, the emergence of $\log$-corrections makes any definite determination
of scaling behavior impossible over the range of system sizes $N$
accessible here, and any of the following scaling relations should
be viewed as purely phenomenological. 

The scaling  becomes increasingly worse with $\sigma>0.5$, signaling
a new $N$ dependence.

We believe this change in behavior at $\sigma=0.5$ is due to the
topological change, from IR to LR, that occurs at this point (see
the discussion in the previous section). Even though each node in
the LR regime is still connected to other spins at arbitrarily large
distances, its possible influence is now limited to a fraction of
the total number of variables  in the system, resulting in
smaller avalanches. The avalanche size effectively creates a limit
on the length of the jumps in configuration space that the system
is capable of performing, forcing a less than optimal sampling of
phase space, and increasingly poorer results.

Avalanche sizes are determined by the total number of spin flips
that occur. If the same spin happens to flip several times, then it
will be counted multiple times as well, but we can also  count
the number $U$ of just which spins flip at least once. The ratio
$S/U$ of the avalanche size, $S$, over the number of unique spins
flipped, $U$, gives us a measure of how important loops are in the
dynamics of the system, a large ratio will indicate that perturbations
spread throughout the system and keep returning to the same spin,
while a number close to unity would mean that avalanches propagate
in just one direction and never double back. 

\begin{figure}
\includegraphics[clip,width=7cm]{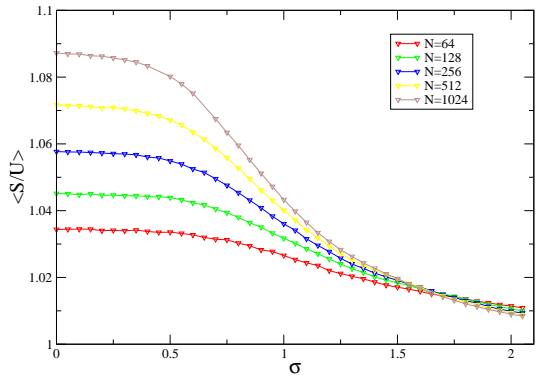}

\caption{\label{fig:unique}Ratio between the avalanche size and the number
of spins that were flipped. }
\end{figure}

On     Fig. \ref{fig:unique} we plot $\left\langle S/U\left(\sigma\right)-1\right\rangle $
for different system  sizes. We find that the ratio between the
size of the avalanche and the number of unique spins flipped is always
small and becomes approximately system size independent in the Short
Range phase. This confirms, once again, that in this region the sphere
of influence of each spin is very small, being limited practically
only to Nearest Neighbors.

\begin{figure}
\includegraphics[clip,width=7cm]{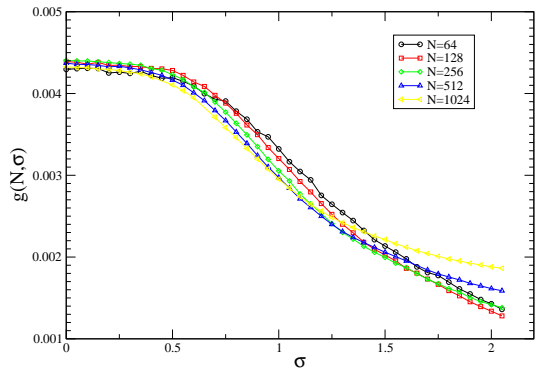}

\caption{\label{fig:unique-scale} Scaling function $g(N,\sigma)$, as defined
in Eq. (\ref{eq:SUscaling}).}
\end{figure}

This quantity obeys a phenomenological scaling relation  of the form:

\begin{equation}
\left\langle \frac{S}{U}\left(N,\sigma\right)-1\right\rangle =\sqrt{N}\left[g\left(N,\sigma\right)-A\log\left(N\right)\right]\label{eq:SUscaling}\end{equation}
where $A\approx4\times10^{-4}$is a small constant and $g\left(N,\sigma\right)$
is shown in Fig. \ref{fig:unique-scale}. The scaling collapse of
the data is very good up to near $\sigma\approx1.5$ where the system
acquires a purely short range behavior. {[}Clearly, this collapse
is purely phenomenological, as the $\log$-correction in Eq. (\ref{eq:SUscaling})
would ultimately overwhelm the constant term.]

Finally, we study how the algorithm approaches the final configuration,
at $H\approx0$ and $m=0$, by looking at the auto-correlation function
given by:\begin{equation}
\langle S_{i}^{0}S_{i}^{\tau}\rangle-\langle S_{i}^{0}\rangle\langle S_{i}^{\tau}\rangle\label{eq:autocorr}\end{equation}
where the indices denote summation over all spins $i$. In Eq. (\ref{eq:autocorr}),
we measure the overlap between the final configuration and those obtained
a number of $\tau$ \emph{complete cycles} backwards in the past at
their $H=0$-crossing. Intuitively, we expect that the configurations
seen at the beginning of the procedure (large values of $\tau$) will
be completely unrelated to the final configuration ($\tau=0$), resulting
in a value near zero for this quantity. However, as the algorithm
takes its course and approaches its conclusion, so too must the configurations
start approaching the final one, corresponding to a value close to
$1$. The way in which it varies from values near $0$ to values close
to $1$ gives us information about the way exploration of configuration
space occurs. The longer the period during which the correlations
are close to $0$, the larger the volume explored, and the faster
it gets close to zero, the earlier the system restricts itself to
a given region, thus limiting the quality of the solution it is able
to find.

\begin{figure}
\includegraphics[width=7cm]{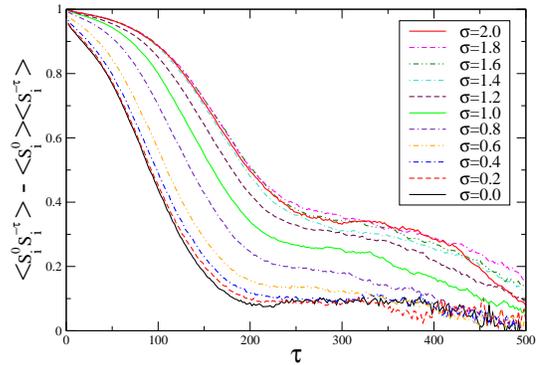}

\caption{\label{fig:corr}Correlation with the final solution as a function
of the time in the past.}
\end{figure}

In Fig. \ref{fig:corr} we plot this quantity for the case of $N=256$,
averaged over $1000$ different instances for each value of $\sigma$
and with $10$ different runs per instance. For small values of $\sigma$,
the plateau at low correlations is extended (lower solid black curve),
followed by an increase towards the value of $1$ near the final stages
$\tau\to0$. As $\sigma$ increases, the auto-correlations increase
within the plateau which itself shortens, and the tendency towards
$1$ becomes noticeable right from the onset (upper solid red
curve for $\sigma=2$). This is a clear demonstration of the ideas
expressed earlier, that the volume of configuration space explored
becomes smaller with the decrease in avalanche size corresponding
to increasing $\sigma$.

\section{Discussion}

The performance of the Hysteretic Optimization procedure for spin
glasses was analyzed for a spin glass model that interpolates between
systems with highly connected variables in the mean field limit and
sparsely connected variables in the nearest-neighbor lattice limit.
$HO$ is shown to be  very fast, but the quality of it's solutions
quickly start decaying for increasing values of the distance dependence
exponent, $\sigma$. An analysis of the avalanche dynamics occurring
in these systems revealed that the failure of $HO$ is due to the
truncation of avalanche size, and hence a limited exploration of the
energy landscape, that occurs when the system is no longer in the
Infinite Range phase. 

The analysis of the behavior of the auto-correlation function with
$\sigma$ confirmed this idea by showing that $HO$ becomes stuck
in a limited region of configuration space increasingly earlier for
larger values of the distance dependence exponent, $\sigma$.

$HO$, being dependent on avalanches for its  local search, cannot
continue to work when the avalanches are no longer large enough to
facilitate large jumps in configuration space. Any attempt to use
$HO$ in a finite connected system, such as an Edwards-Anderson spin
glass or many combinatorial optimization problems\cite{Percus06a},
is, therefore, inefficient (see  Chap. 10 in Ref. \cite{Dagstuhl04}
and Ref. \cite{Zapperi04}). Our attempts to simulate sparsely connected
systems, in this case 3SAT\cite{Percus06a} and EA spin glasses with
$\pm J-$bonds, with \emph{discrete} bond weights proved particularly
unsuccessful. In such system, all variables only possess a finite
(and typically, small) range of local field states to take on. For
instance, in such an EA spin glass in $d=3$ dimensions, all spins
have exactly $2d+1=7$ states. Thus, a hysteresis loop has just 7
jumps between full up- and and full-down saturation. At each jump,
a finite fraction of spins flip simultaneously due to degeneracies,
but mostly in an \emph{uncorrelated} manner dictated by their local
environment. An open question remains that concerns problems defined
on random graphs of finite connectivity but with a continuous distribution
of bond-weights, such as the Viana-Bray spin glass\cite{Viana85}
with a Gaussian bond distribution. Unlike for a lattice, the number
of neighbors increases exponentially with distance so that every node
is connected to every other node with $\sim\ln(N)$ steps, and even
small correlations could quickly span the system. One might expect
that there would be a crossover between the average connectivity and
$\ln(N)$ separating broadly distributed avalanches from localized
ones, which would be very weak. Hence, HO might still work reasonably
well in those systems down to low connectivities for most practical
system sizes. In fact, our preliminary studies of Gaussian spin glasses
on 3-connected graphs showed only minor deterioration in HO compared
to EO for increasing system sizes (up to $N=1023$). Yet, an independent
comparison of HO to itself within a one-parameter family of models
in the spirit of our approach here would require much more simulation
for variable connectivity.

These results also highlight one important ingredient for any efficient
algorithm or heuristic: the ability to travel between very distant
regions of configuration space without being impaired by the large
energy barriers that make such jumps energetically or entropically
unfavorable. This ability is only within $HO$s reach for Infinitely
Range systems.

\section{Acknowledgments}

We would like to thank Helmut Katzgraber for inspiring discussions
and the Division of Materials Research at the NSF for their support
under grant \#0312150 and the Emory University Research Council for
seed funding. 

\bibliographystyle{unsrt}
\bibliography{papers,spinglass}

\end{document}